\documentclass[amsmath,amssymb,prl,twocolumn,showpacs,final]{revtex4}

\usepackage{amsmath}
\usepackage[applemac]{inputenc}			

\usepackage{graphicx}
\usepackage{amssymb}
\usepackage{bm}
\newcommand{\be}{\begin{equation}} 
\newcommand{\ee}{\end{equation}}

\usepackage{color}
\usepackage{li psum}

\begin{document}


\title{Effective Wall Friction in Wall-Bounded 3D Dense Granular Flows}

\author{Riccardo Artoni}
\email{riccardo.artoni@ifsttar.fr}
\author{Patrick Richard}

\affiliation{LUNAM Universit\'e, IFSTTAR, MAST, GPEM, F-44340 Bouguenais, France.}%
\date{\today}

\begin{abstract}
We report numerical simulations on  granular shear flows confined between two flat but frictional sidewalls. Novel regimes differing by their strain localization features are observed. They originate from the competition between dissipation at the sidewalls and dissipation in the bulk of the flow. The effective friction at sidewalls is characterized (effective friction coefficient and orientation of the friction force) for each regime, and its interdependence with slip and force fluctuations is pointed out. 
We propose a simple
scaling law linking the slip velocity to the granular temperature  in the main flow direction which leads naturally to another scaling law for the effective friction. 
\end{abstract}
                              

\maketitle
Recent progress has been made in the theoretical description of the rheology of granular materials~(e.g. \cite{kamrin12,Kamrin_CompPartMech_2014,Henann_PNAS_2013,Kamrin_SoftMatter_2015,Henann_PRL_2014,Chialvo_PoF_2013,Bouzid_PRL_2013,Bouzid_EPL_2015}). However, a scientific bottleneck still prevents the use of those theories in real configurations: the modeling of the interaction of the granular material with a solid boundary 
and its implication on the rheology of the system~\cite{Staron_EPJE_2014,Artoni_ChemEngSc_2011}. 
Yet, such interactions are crucial to understand shear banding that is a widespread phenomenon in
slow granular flows.
They are also essential for the comprehension of many industrial and agricultural applications (silos, hoppers, chutes, mixers, and blenders) and geophysical phenomena, for example,  the mobility of avalanches or the cratering due to an impact.\\
The aforementioned difficulty originates from experimental and numerical evidence that questions the modeling of a solid interface as a simple boundary condition. Indeed,
the existence of cooperative effects~\cite{Pouliquen_PRL_2004,Staron_PRE_2008} in the force network and in the velocity field prevents the use of a purely local approach, {i.e.}, an approach in which 
 the local
stresses are simply related to the local shear rate. 
An important source of cooperative effects and thus of nonlocality is the mechanical noise of the flow itself~\cite{behringer2008,nichol10,reddy11,Bouzid_EPL_2015}.
Thus, the  effective friction coefficient  of a granular material flowing on a flat but frictional interface and the corresponding slip velocity are partially controlled by the shear-induced fluctuations of the force network~\cite{Richard_PRL_2008, artoni09, artoni12}, and their prediction remains a challenge. 
The understanding of such interfacial phenomena is of paramount importance not only for granular flows but also for other complex fluids displaying wall slip and nonlocal effects such as emulsions~\cite{Mansard_SM_2014}, dense suspensions~\cite{ballesta2012,gibaud08}, foams, etc.
Here, by means of discrete element simulations, we study the effective friction of a dense granular material confined in a shear cell. We show that at flat frictional walls, even if the system is globally below the slip threshold, force fluctuations trigger slip , leading to a nonzero slip velocity and an effective wall friction scaling with a sliding parameter. These results shed light on the necessity to introduce fluctuations in theories aiming to capture and predict the behavior of dense granular flows at the vicinity of an interface. 

\begin{figure}[!ht]
\includegraphics[width=\columnwidth]{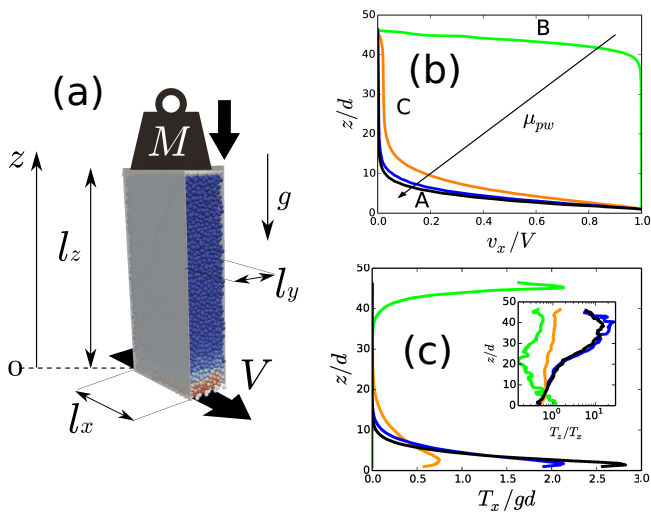}
    \caption{(color online) (a) Sketch of the wall-bounded flow configuration. (b) Velocity profiles at the sidewalls along $z$ obtained for $\tilde V =10 $, $\tilde M=0.2$, and different values of the particle-wall friction coefficient: $\mu_{\rm{pw}}= 0.05,\ 0.1,\ 0.2$ and $0.3$. Three patterns of localization appear.  (c) Fluctuation of the $x$ particle velocities at the wall along $z$ for the same set of parameters. Inset: Ratio of vertical to horizontal fluctuations versus $z$.}\label{figure1}
\end{figure}

\paragraph{Numerical simulations.---}We carry out and analyze numerical simulations of dense granular flows in a 3D wall-bounded geometry. Simulations are performed by using the contact dynamics method \cite{jean99}, as implemented in the LMGC90 open source framework \cite{renouf04}. The flow configuration [sketched in Fig.~\ref{figure1}(a)] is a rectangular cuboid (length $l_x=20d$, depth $l_y=10d$, and variable height $l_z$) characterized by a periodic boundary condition in the main flow direction ($x$), two bumpy walls at the top and at the bottom, and two lateral flat but frictional walls (normal to the $y$ direction).  Gravity acts on the system along $z$, and the flow is driven by the motion along $x$ of the bottom wall.  The top wall cannot move on the $x$ and $y$ directions but is free to move in the $z$ direction, simply according to the balance between its weight and the force exerted by the grains. Simulations were performed with $N=10000$ slightly polydisperse spheres (uniform number distribution in the range $0.9 d-1.1 d$)  interacting through perfectly inelastic collisions and Coulomb friction ($\mu=0.5$). 
The coefficient of restitution has nearly no influence on dense granular flows due to the presence of enduring contacts~\cite{dippel1999}. Consequently, we chose perfectly inelastic grains to maximize dissipation and thus save computation time.
Interactions of particles with the flat walls were also perfectly inelastic and frictional (with a coefficient of friction $\mu_{\rm{pw}}$). We performed several simulations varying the velocity of the bottom wall $V$, the weight of the upper wall $M$, and the particle-wall friction coefficient $\mu_{\rm{pw}}$. The first two parameters can be made dimensionless, for example, by considering a particle Froude number $\tilde V ={V}/{\sqrt{gd}}$ and the ratio between the mass of the top wall and the mass of the grains, $\tilde M= {M}/{N m}$, where $m$ is the average particle mass.
This flow configuration is interesting for studying wall friction, since slip velocity and fluctuations are less dependent on the wall properties than if shear acted on planes parallel to the wall. This does not mean that the wall properties do not affect velocity profiles, but that the problem is a little more conceptually decoupled than, for example, in inclined chute flows.
Preliminary analysis indeed showed that velocity and velocity fluctuation profiles were nearly uniform in the $y$ direction for the range of parameters considered in this study  ($\tilde V= 0.1 - 10$, $\tilde M= 0.2 -2$, $\mu_{\rm{pw}}=0.05-0.3$). The profiles at the wall can therefore be taken also as a reference for the internal behavior; shear principally acts on a plane orthogonal to the wall. It is clear that not all variables are uniform in $y$: for example, solid fraction $\phi$ is influenced by the presence of walls and displays fluctuations around a decreasing value when approaching the wall~\cite{Camenen_PRE_2012}. On the other hand, due to wall friction, stress components $\sigma_{yz}$ and $\sigma_{yx}$ (and their symmetric counterparts) will vary with $y$. In this work, we focus mainly on wall behavior, postponing a full analysis of the behavior of the system including also profiles in the $y$ direction to a more detailed study.

We computed average profiles along $z$ by performing averages on slices with a thickness  $2d$ in the vertical direction. 
The streamwise effective wall friction coefficient at the lateral walls is estimated for each slice as the ratio of the average force in the flow direction $x$ and the average force in the direction normal to the wall $y$: $\mu_w={\langle F_x\rangle}/{\langle F_y \rangle}$~\cite{Richard_PRL_2008}. This corresponds to the stress ratio $\sigma_{yx}/\sigma_{yy}$ at the wall. Profiles of velocity fluctuations in the $x$ and $z$ direction (which are related to granular temperature) are also calculated as $T_{x}=\langle(v_x -\langle v_x \rangle )^2\rangle $, $T_{z}=\langle (v_z -\langle v_z\rangle)^2\rangle$, where the fluctuations are correctly computed with respect to the average velocity value extrapolated at the particle center \cite{artoni15}. 
Concerning bulk profiles, we computed for each slice the average value of  solid fraction $\phi$ (not discussed here) and stress components $\sigma_{xx}$, $\sigma_{yy}$, $\sigma_{zz}$, and $\sigma_{xz}$. The latter allow to define a bulk friction coefficient $\mu_{xz}=\sigma_{xz}/p$	 where $p=(\sigma_{xx}+\sigma_{yy}+\sigma_{zz})/3$.

\paragraph{Flow profiles.---}An example of the velocity profiles obtained in this configuration is given in Fig.~\ref{figure1}(b), for $\tilde V =10 $, $\tilde M=0.2$, and different values of the particle-wall friction coefficient. Profiles typically display shear localization. For the range of parameters considered in this study, three regimes appear: (A) for high $\tilde M$ and/or high wall friction, shear is localized at the bottom; (B) for low $\tilde M$ and low wall friction, shear is localized near the top; and (C) for low $\tilde M$ and intermediate wall friction, a central plug can form with two shear zones near the bumpy walls.  Concerning such profiles, in the following we will refer to the ``shear zone'' as the zone where most of the velocity variation occurs (say, 99\%) and to the ``creep zone'' as the nearly stationary remaining part. For regimes B and C, far from the shear band, it is better to refer to a ``plug flow zone.'' In shear zones, velocity profiles are characterized by an exponential variation whose characteristic length is mainly a function of $\mu_{\rm{pw}}$ and $\tilde M$. If $\tilde M \gg 1$, bottom localization seems to be more probable and  the characteristic length decreases with $\mu_{\rm{pw}}$; it is reasonable to infer that a linear profile should be obtained for sufficiently large $\tilde M$ and sufficiently low $ \mu_{\rm{pw}}$. Such a coexistence of different localization patterns was recently reported for a different flow configuration~\cite{moosavi2013}. Figure~\ref{figure1}(c) displays sample profiles of streamwise velocity fluctuations $T_x$. These agree qualitatively with shear rate profiles, with higher fluctuations in the shear zones. Vertical fluctuations, as shown in the inset in Fig.~\ref{figure1}(c), tend to be more important than streamwise ones in the creep zone in regime A.
In the shear zones, there was also evidence of the formation of secondary (weak) convective rolls, similar to those observed in Refs.~\cite{brodu13,Brodu_JFM_2015}. These rolls appeared to be confined to the shear zone, to be stronger for regime C (where the shear zone was larger), and were characterized by particles moving upwards near the flat walls and descending in the center of the channel. 
The localization pattern observed, as well as the convective instability, is very new and interesting and will be addressed in a detailed study.

\begin{figure}[ht]
\includegraphics[width=\columnwidth]{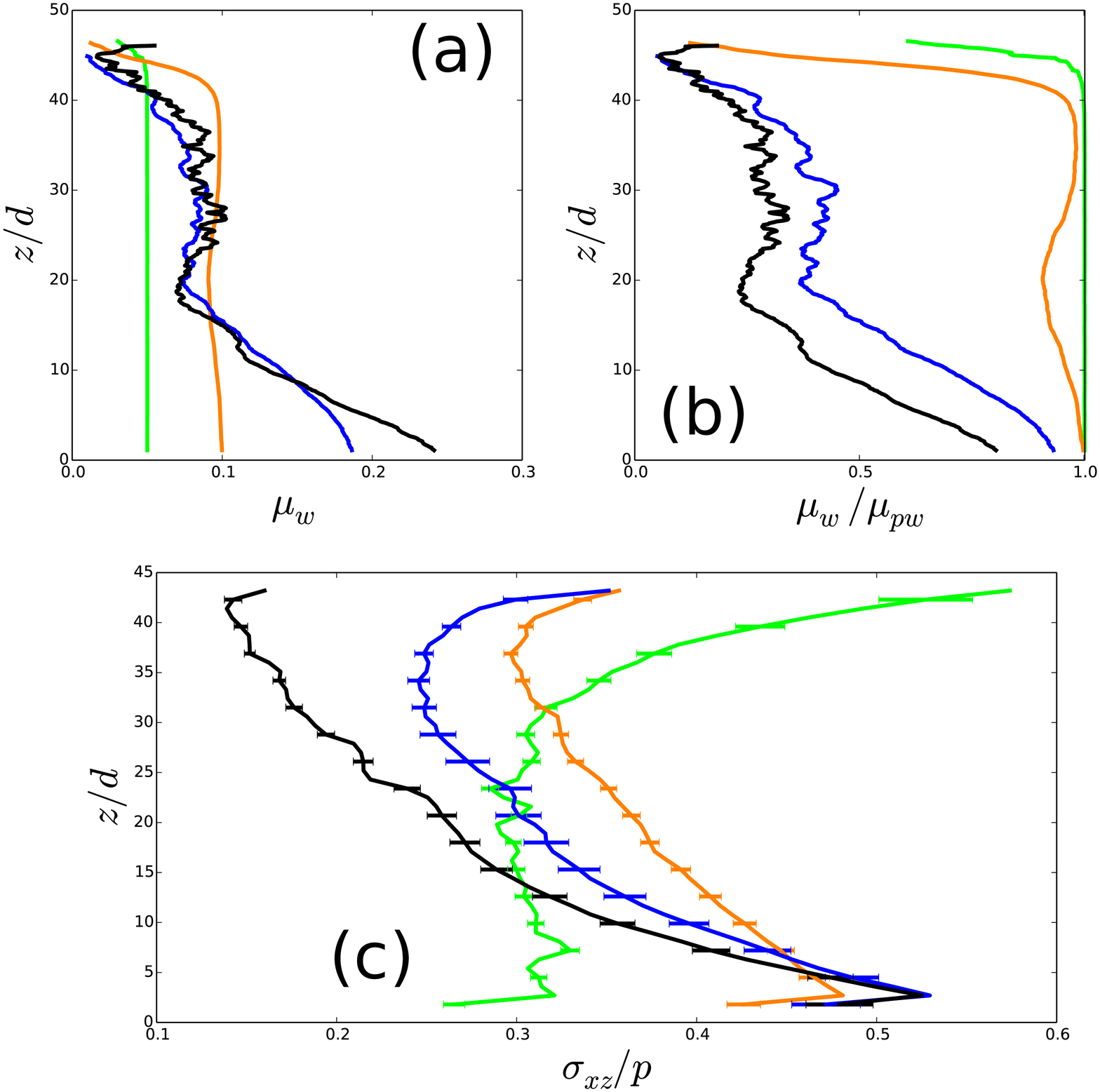}
    \caption{(color online) Profiles along $z$ of  (a) the effective wall friction coefficient $\mu_w$ and  of (b) $\mu_w$ rescaled on the particle-wall friction coefficient $\mu_{\rm{pw}}$,  for  $\tilde V =10 $, $\tilde M=0.2$, and different values of $\mu_{\rm{pw}}= 0.05,0.1,0.2,0.3$. $\mu_w/\mu_{\rm{pw}}$ decreases with $\mu_{\rm{pw}}$. (c) Profiles of the bulk friction coefficient $\mu_{xz}=\sigma_{xz}/p$ along $z$. The labels are the same as those used in Fig.~\ref{figure1}(b).
}\label{figure2}
\end{figure}
\paragraph{Effective wall friction.---} Figures~\ref{figure2}(a) and \ref{figure2}(b) display profiles of the effective wall friction coefficient, for $\tilde V =10 $, $\tilde M=0.2$, and different values of $\mu_{\rm{pw}}$. It is clear that, as  was seen for 2D flows \cite{artoni12} or in 3D confined gravity-driven flows~\cite{Richard}, $\mu_w/\mu_{\rm{pw}}$ decreases with $\mu_{\rm{pw}}$. These profiles are strongly related to velocity profiles: in regime B, far from the shear band, grains move as a plug in the $x$ direction with velocity $V$. There are no stick events, so the effective coefficient of  friction in the plug zone is equal to $\mu_{\rm{pw}}$. In the shear zone, stick-slip events may emerge, and therefore $\mu_w < \mu_{\rm{pw}}$~\cite{Richard_PRL_2008}. For the simulations where shear is localized at the bottom (regime A), we find a decreasing profile of $\mu_w$ versus $z$ in the shear band and a less pronounced variation in the creep zone. This can be explained by the fact that, reasonably, stick-slip events become more and more probable when we approach the creep zone~\cite{Richard_PRL_2008}. When a central plug forms (regime C), a more complex $\mu_w$ profile may appear, displaying a flat local maximum corresponding to the plug zone. These results confirm what was already known for 2D flows, that is, the interdependence of effective friction, slip and fluctuations \cite{artoni09, artoni12}. 
The effective bulk friction coefficient $\mu_{xz}$ as shown in Fig.~\ref{figure2}(c) qualitatively follows the behavior of velocity fluctuations and shear rate. An increase in wall friction contributes to weakening of the effective bulk friction in the creep zone. The effective bulk friction coefficient  being generally higher than the effective wall friction coefficient, in the creep zone they can take similar values.

\begin{figure}[t!]
\includegraphics[width=\columnwidth]{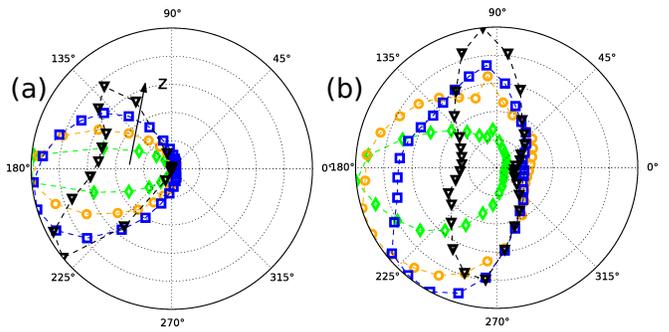}
    \caption{(color online) Polar representation of the probability density function of tangential force orientation at the wall,  for  $\tilde V =1 $, $\tilde M=0.2$, at several heights $z$ ($=3d$, $9d$, $18d$, and $36d$), for (a) $\mu_{\rm{pw}}=0.1$ (regime C) and (b) $\mu_{\rm{pw}}=0.3$ (regime A).  Statistics are performed on a $5d$-wide slice centered on the given $z$ value.}\label{figure3}
\end{figure}

\paragraph{Force orientation at the wall.---} The probability density function of tangential force orientation depends on the flow regime and flow zone. For regime B, in the plug flow zone, tangential forces are purely oriented along $x$ (counterflow), and in the shear zone only slight differences exist. For regimes A and C, in the shear zone [see, for example, curves at $z=3d$ in Figs.~\ref{figure3}(a) and~\ref{figure3}(b)], the most likely direction is again the $x$ direction, but significant vertical fluctuations are present. 
The likelihood of such transversal fluctuations increases with the wall friction.
For these two regimes, the probability of finding vertical force fluctuations is more important for grains close to the top stationary wall. 
In the creep flow zone of regime A [curve at $z=36d$ in Fig.~\ref{figure3}(b)], we see that vertical force orientations can become more probable than horizontal ones. This is obviously related to the incidence of vertical velocity fluctuations discussed previously. The average tangential force is always directed opposite to the motion, but a very wide distribution of orientations is present. The probability distribution function (PDF) is roughly ellipsoidal. For regime C in the plug flow zone,  due to the partial slip of the plug,  the PDF is similar to the creep zone PDF  but is bent towards the $x$ direction, assuming  a sort of ``bunny ears'' shape.
Note that the reported behavior is similar to that obtained for confined granular flows on a heap, where a modification of the orientation of the friction force which is significant in the creeping zone has been reported~\cite{Richard_PRL_2008}.

\begin{figure}
\includegraphics[width=\columnwidth]{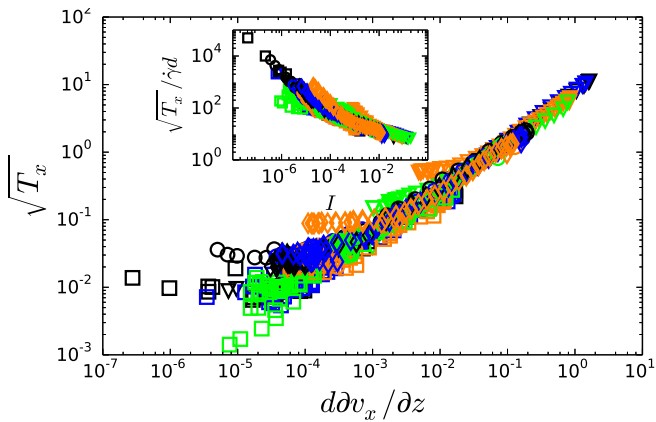}
\caption{(color online) Comparison between a velocity scale based on shear rate,  $d \partial v_x /\partial z$, and one based on $x$-velocity fluctuations, $\sqrt{T_x}$ for $\tilde{M}=0.2$ and $\tilde{V}=0.1$ (squares), $\tilde{M}=0.2$ and $\tilde{V}=1$ (diamonds), and $\tilde{M}=0.2$ and $\tilde{V}=10$ (downward triangles), $\tilde{M}=2$ and $\tilde{V}=1$ (circles). The colors (gray scales) correspond to different grain-wall friction coefficients [see Fig.~\ref{figure1}(b)]. The fluctuational velocity scale displays a plateau for each simulation for low shear rates. Inset: $\sqrt{T_x}/\dot\gamma d $ increases when decreasing inertial number $I$. The labels are the same as those used in Fig.~\ref{figure1}(b).}\label{figure4}
\end{figure}

\paragraph{Wall friction scaling.---} In order to study more in detail effective wall friction, we discuss now its scaling on system parameters. For 2D dense flows on an incline, it was shown that  $\mu_w/\mu_{\rm{pw}}$ scales with a dimensionless slip length defined as $v_{\rm{slip}}/\dot \gamma d$ \cite{artoni12}. The origin of this scaling was postulated to be the presence of force and velocity fluctuations, yielding stick-slip events. The inverse of the shear rate $\dot\gamma^{-1}$ was there used as an estimate of the time scale of force fluctuations. In the 3D flow configuration discussed here, this scaling has some drawbacks  and seems not to 
hold well in the creep zone. One reason for this behavior is that $\dot\gamma d$ may not be a good velocity scale for slip events in creeping flows, as seems to be the case for dense flows. Another estimate  for a velocity scale related to force fluctuations may be the correlation of velocity fluctuations, e.g., granular temperature. The latter quantity
was indeed used as a velocity scale in boundary conditions for flat frictional walls in the kinetic theory of granular gases \cite{johnson_jackson_1987,richman88,jenkins_1992}. 
Our simulations show that the   square root of the velocity fluctuations in the $x$ direction ($\sqrt{T_x}$)  globally scales with the shear rate (see Fig.~\ref{figure4}).
However, often in the creep or plug zone, when $\dot\gamma$ decreases, the fluctuations decrease less rapidly, eventually displaying a plateau. This is evident from the inset in Fig.~\ref{figure4}, where $\sqrt{T_x}/\dot\gamma d $ is shown to increase when decreasing the inertial number $I=\dot\gamma d / \sqrt{p/\rho}$ (creep is found for $I\rightarrow 0$).  The reason for this plateau can be related to the nonlocality of the rheology: it was already shown that fluctuations propagate far from the shear zone \cite{bocquet02,nichol10,reddy11}.  Such nonlocal effects can be modeled by the diffusion of fluctuating energy, as in kinetic theories of dense granular flows \cite{JenkinsBerzi2010,artoni11}, or by fluidity-based models \cite{kamrin12}.  Note that the plateaus are even more evident for vertical velocity fluctuations  ($\sqrt{T_z}$, not shown), which are probably influenced by vertical fluctuations of the position of the upper wall. 
 
\begin{figure}
\includegraphics[width=\columnwidth]{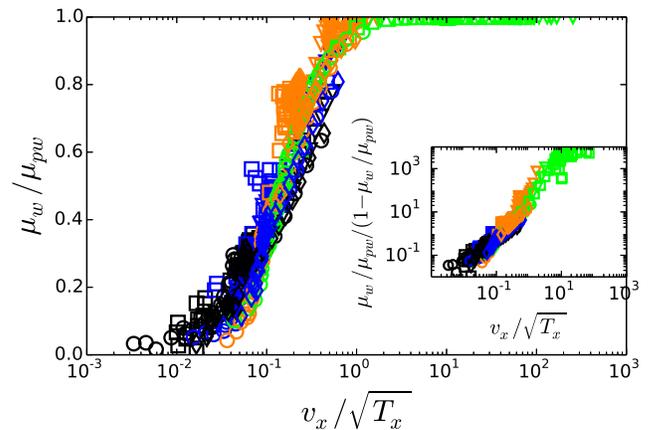}
    \caption{(color online) Rescaled effective wall friction coefficient versus the dimensionless slip parameter $v_x/\sqrt{T_x}$. The inset shows that a power law relationship holds between $\frac{\mu_w/\mu_{\rm{pw}}}{1-\mu_w/\mu_{\rm{pw}}}$ and $v_x/\sqrt{T_x}$.   
		The labels are the same as those used in Fig.~\ref{figure4}.}\label{figure5}
\end{figure}

The correspondence between the plateau of the curves $\sqrt{T_x}$ versus $d\partial v_x / \partial z$ and the $\mu_w$ profile in the creep zone suggests to test the $\sqrt{T_x}$ scaling for the slip velocity.
This test is shown in Fig.~\ref{figure5}, where we display the rescaled effective friction coefficient $\mu_w/\mu_{\rm{pw}}$  as a function of $v_x/\sqrt{T_x}$. Note that it is preferable to use the streamwise velocity scale $\sqrt{T_x}$ (and neither $\sqrt{T_z}$ nor $\sqrt{T_x + T_z}$) given that also $\mu_w$ is based on the forces oriented along the main flow direction.  
Note also that $T_y$ might be a better choice for more dilute flows where most grains interacting with sidewalls bounce on them. In the present study, flows are dense, and grains mostly slide on sidewalls.
We can see that the scaling performs globally well on several orders of magnitude. There are indeed some deviations from the master curve, in particular, some simulations in regime C (orange squares, diamonds, and triangles in  Fig.~\ref{figure5}) which show slightly higher values of $\mu_w/\mu_{\rm{pw}}$ than the other simulations.
This could be related to the presence of convective structures discussed previously. This kind of structure is known to affect effective friction \cite{brodu13,Brodu_JFM_2015} and may be the cause of the slight discrepancy between these simulations and the rest of the set.
The functional form of the scaling appears to be the same as in Ref.~\cite{artoni12} (with the exception of the different velocity scale): $$\frac{\mu_w}{\mu_{\rm{pw}}}=\frac{ \left({v_x}/{\sqrt{T_x}}\right)^B}{A+\left({v_x}/{\sqrt{T_x}}\right)^B}.$$ For the present case, we estimated $A= 1.2$ and $B= 1.7$ by curve fitting. 
It should be pointed out that this scaling is similar to the boundary conditions of kinetic theories of granular flows~\cite{johnson_jackson_1987,richman88,jenkins_1992} and used in the extension of the kinetic theory proposed by Jenkins and Berzi~\cite{JenkinsBerzi2010,Jenkins_GranularMatter_2012}. Moreover, a connection also exists with the nonlocal theories recently developed~\cite{kamrin12,Henann_PNAS_2013,Kamrin_CompPartMech_2014,Henann_PRL_2014,Kamrin_SoftMatter_2015}, since the granular fluidity, i.e., the ratio  of the pressure to the viscosity, can be shown to scale with the square root of the granular temperature.

\paragraph{Conclusions.---} Results presented in this work confirm for 3D flows the conceptual framework which was suggested in Ref.~\cite{artoni09}; at flat frictional walls, force fluctuations trigger slip events even if the system is globally below the slip threshold. These stick-slip events produce (i) a nonzero average slip velocity and (ii) a variable effective wall friction coefficient which scales on a dimensionless slip parameter. It is interesting to see that this picture of wall slip (fluctuations leading to a finite probability of yielding) is conceptually the same as that at the heart of recent findings on granular creep flows~\cite{nichol10,reddy11}.
The scaling of the slip velocity on the rms velocity fluctuations supports also the idea that fluctuations are a key ingredient that should be included in any model aiming to describe dense granular flows at the vicinity of an interface.
The scaling law can be used as a boundary condition in a nonlocal theory for dense granular flow implying granular temperature~\cite{JenkinsBerzi2010,Jenkins_GranularMatter_2012}. In order to solve such a model, an additional constraint is needed: a possible choice is to equate the flux of fluctuating energy at the boundary to the dissipation, as in kinetic theories. An expression for the energy dissipation based on frictional sliding should be studied.
In Ref.~\cite{artoni12}, a scaling based on the shear rate was shown to be effective for 2D inclined chute flows. In that case, given that the shear acted in planes parallel to the wall,  the scaling law could be interpreted as a Navier partial slip law. The present contribution shows that such a partial slip boundary condition involving the velocity gradient normal to the wall cannot be universal. Fluctuations trigger slip events and therefore affect effective friction, whether shear planes are parallel or not to the wall. Thus, if a scaling on shear rate was valid in that case, it was probably because the shear rate was a good measure of the scale of fluctuations for the considered data.\\
Understanding the behavior of granular materials at a flat but frictional interface is a preeminent scientific challenge. In this work, we demonstrated the relevant role of fluctuations and pushed  forward our understanding of effective wall friction and wall slip. In the future, it will be interesting to test the framework relating fluctuations, wall friction and wall slip in other contexts, the problem of interface rheology being common to other divided media, such as foams, clays, or gels. The effect of grain shape should also be explored, since it may influence the wall-slip velocity and thus the reported scalings.

\begin{acknowledgments}
The numerical simulations were carried out at the CCIPL (Centre de Calcul Intensif des Pays de la Loire) under the project
MTEEGD.
\end{acknowledgments}

\end{document}